# A Bayesian Committee Machine Potential for Oxygen-containing Organic Compounds

Seungwon Kim,[1] D. ChangMo Yang,[1] Soohaeng Yoo Willow,[1, *] and Chang Woo Myung[1, †]

[1]*Department of Energy Science, Sungkyunkwan University, Seobu-ro 2066, Suwon, 16419, Korea*

(Dated: March 2, 2024)

Understanding the pivotal role of oxygen-containing organic compounds in serving as an energy source for living organisms and contributing to protein formation is crucial in the field of biochemistry. This study addresses the challenge of comprehending protein-protein interactions (PPI) and developing predicitive models for proteins and organic compounds, with a specific focus on quantifying their binding affinity. Here, we introduce the active Bayesian Committee Machine (BCM) potential, specifically designed to predict oxygen-containing organic compounds within eight groups of CHO. The BCM potential adopts a committee-based approach to tackle scalability issues associated with kernel regressors, particularly when dealing with large datasets. Its adaptable structure allows for efficient and cost-effective expansion, maintaing both transferability and scalability. Through systematic benchmarking, we position the sparse BCM potential as a promising contender in the pursuit of a universal machine learning potential.

## INTRODUCTION

Organic compounds containing oxygen atoms are used as an energy source for all living organisms, and they also form proteins by combining them with other atoms. For proteins, protein-protein interaction (PPI) constitutes a cornerstone in various cellular processes, encompassing signal transduction pathways, metabolic regulation, and cell cycle progression [1–3]. The binding affinity between interacting proteins fundamentally governs the specificity and effectiveness of PPI [4].

Quantifying the binding affinity for PPI often involves equilibrium dissociation constant ($K_d$) and Gibbs free energy ($\Delta G$) [5]. While some protein-protein complexes possess binding affinity data along with their three-dimensional structure, enabling the development of structure-based prediction methods, the experimental procedures for measuring $K_d$ remain labor-intensive and time-consuming, hindering their application in high-throughput affinity estimation [6–8]. Consequently, computational tools leveraging sequence information have been introduced, but they encounter limitations, such as small datasets, performance degradation, and challenges in handling multiple-reaction scenarios involving atom-atom bindings [9].

With the continuous expansion of available data and advancements in machine learning technology, there is an ongoing opportunity to enhance methods for a more efficient and accurate analysis of PPI networks and their underlying mechanisms [10]. This necessity has driven the exploration of new prediction methods capable of effectively capturing key descriptors to predict the binding affinity of protein-protein complexes with diverse functionalities [11].

One noteworthy approach involves the use of the Force Field method [12–17], which delves into the energy relationships between atoms to understand intricate protein-protein interactions [18]. While Force Field methods reproduce some experimental data well, they exhibit shortcomings in explaining helix structures and some intermolecular interactions, challenging the simultaneous and accurate description of correct structures and all properties [19].

Over the past decade, machine learning-based interatomic potentials (MLPs) have emerged as a promising avenue to provide accurate non-parametric representations of potential energy surfaces [20–29]. Various algorithms such as Gaussian approximation potentials [30–35], gradient-domain machine learning [36], Bayesian linear regression variations [37, 38], and sparse Gaussian process regression (SGPR) [39–45], have enabled on-the-fly generation of MLP through molecular dynamics (MD) simulations. However, the development of comprehensive descriptors invariant under transformations and rotations remains a fundamental challenge, affecting the accuracy of MLP.

MD simulations serve as a powerful tool for detailed atomic-level understanding, complementing experimental results. The accuracy of the Potential Energy Surface (PES) can be assessed through comparisons of MD simulation results with experimental observations, such as phase diagrams and radial distribution functions [45]. Improving stress estimation accuracy in MLP [22, 46] can enhance the reliability and efficiency of MD simulations.

To address challenges associated with the SGPR algorithm [39–45], particularly in handling large training datasets and induced local correlation energy (LCE) sets, we try to handle a novel approach using Bayesian Committee Machine (BCM) [46–48]. By employing separately trained MLPs for different subsets and integrating them through BCM, we create a general MLP that not only streamlines the training process but also overcomes computational challenges associated with inverting large kernel matrices (FIG. 1). This approach paves the way for an efficient and accurate general MLP, contributing to the advancement of MLPs in MD simulations.

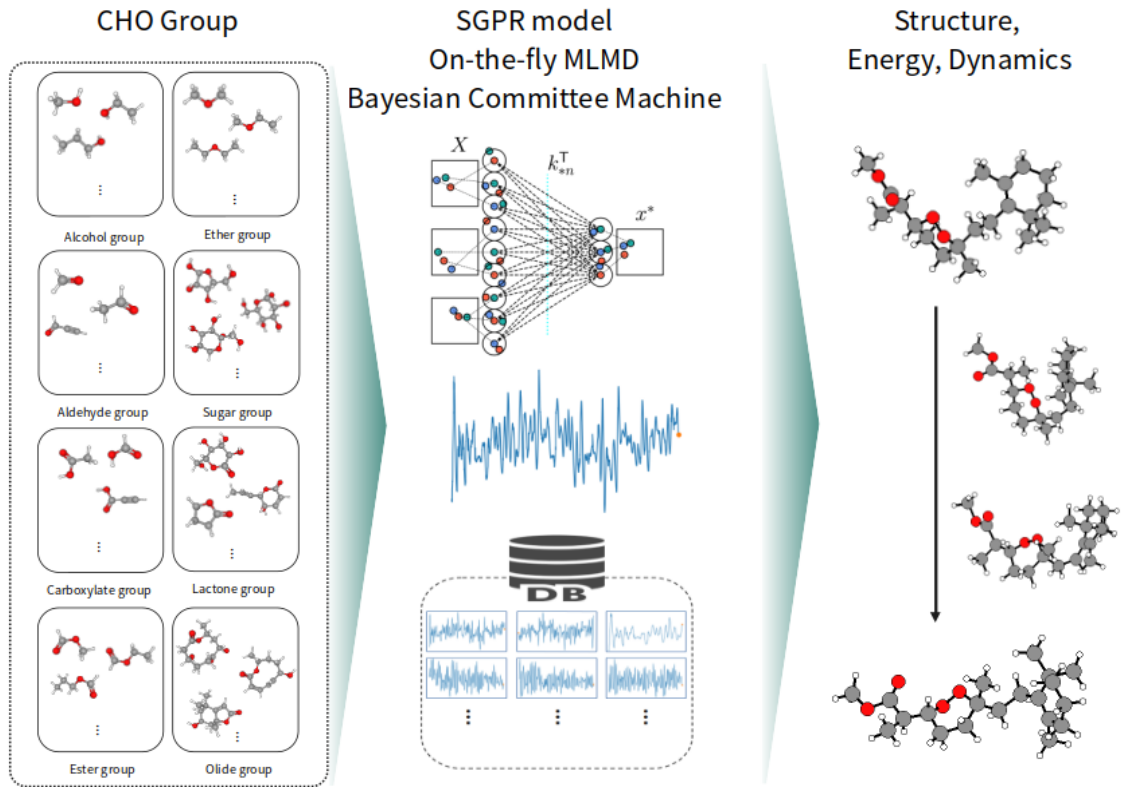

FIG. 1. Schematic of Bayesian committee machine potential where $P$ committee models (sparse local experts in this work) make a final decision on any properties of system. The weight ensures to have confident predictions to be more outspoken than less confident models.

## MATERIALS AND METHODS

Eight groups, distinguished by CHO atoms and classified as alcohol, aldehyde, carboxyl, ester, ether, sugar, lactone, and olide groups, were selected from the ChEMBEL database[49, 50]. Expert models were also trained to count inter-molecular interactions for various molecules featuring characteristics such as arbitrary number of double bonds (C=C and C=O), cyclostructures, aromatic rings, aromatic polycycles, and more within each group (Table S1). The objective was to verify if energy estimation could be achieved from molecular structures with intricate compositions. To obtain a machine learning potential $E_\alpha$ of each expert model, we performed the canonical ensemble (NVT) molecular dynamics simulations using a Nose-Hoover thermodstat and Parrinello-Rahman dynamics, implemented in the atomic simulation environment (ASE) package [51–53].

The MD simulations ran for 3 ps at 300 K with a 0.25 fs time step. The reference potential energy surfaces are from density functional theory (DFT) calculations, performed using the Vienna ab initio simulation package (VASP) [54, 55] with Perdew-Burke-Ernzerhof (PBE) functionals[56] and van der Waals corrections (D3) [57]. We applied projector augmented wave[58] pseudopotentials with a 600 eV energy cut-off, setting the convergence criterion for the electronic energy difference at $10^{-4}$ eV. The molecules were placed in the center of a cubic cell with a 15 Å vacuum interlayer distance, and the Brillouin zone was sampled using the $\Gamma$-point. The local SGPR potential[39] was trained using AUTOFORCE package.[59] And we made the group of 8 MLPs using Bayesian Committee Machine[46, 48]. To check whether the BCM was working or not, root mean square error (RMSE) of per-atom energy and axis-wise RMSE of force for already made a group of 8 MLPs using trajectories for each group.

## RESULTS

The expert SGPR model for each molecular category was set with 300K and 3 ps, and instantaneously generated through short MD simulations of relevant molecules.

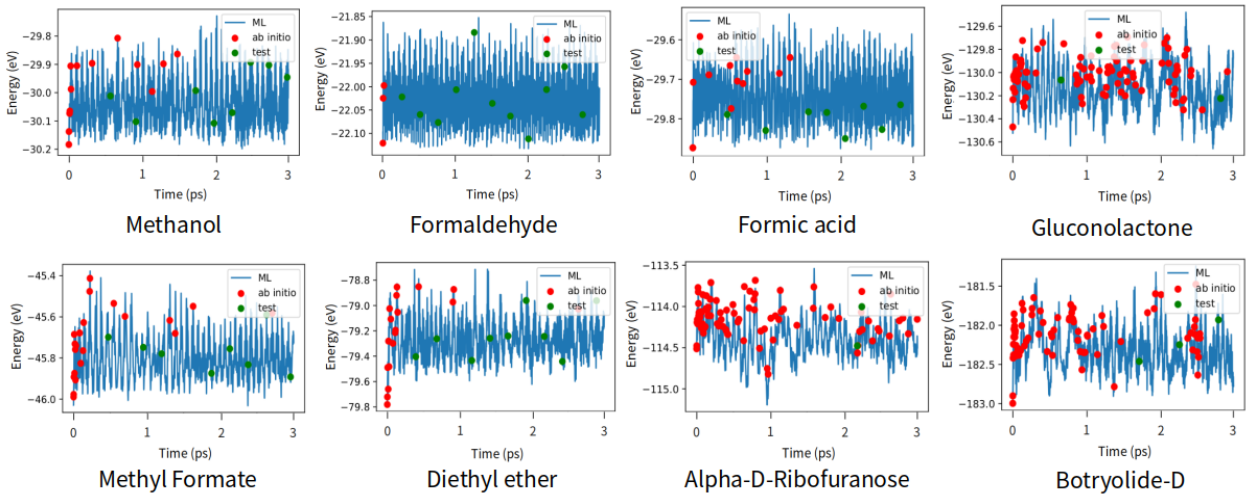

FIG. 2. Initial steps for on-the-fly MLMD for elementary 8 group of molecules. Red bullets indicate the ab initio potential energy of the configurations sampled by the SGPR model. Green bullets are for single-point ab initio tests.

TABLE I. Testing the Expert Models in the 8 groups.

| Group | T [a] | Force MAE for single atom (meV/A) | Potential MAE for single atom (meV) | Force $R^2$ |
|---|---|---|---|---|
| Alcohol | 115 | 8.12106 | 8.66732 | 0.98780 |
| Aldehyde | 76 | 8.40160 | 5.24945 | 0.99210 |
| Carboxylate | 39 | 7.79144 | 3.65209 | 0.99040 |
| Ester | 88 | 5.72440 | 4.43713 | 0.98880 |
| Ether | 62 | 6.53042 | 7.72036 | 0.98800 |
| Sugar | 52 | 6.99712 | 10.45848 | 0.99150 |
| Lactone | 17 | 6.63023 | 4.38984 | 0.99200 |
| Olide | 31 | 5.46338 | 6.83928 | 0.99110 |

[a] The number of sample to train for using POSCAR

For instance, the training for each group potential is illustrated in FIG. 2. The model trained after MD simulations for each molecular group is then used as the initial model for MD simulations of the subsequent molecule. In this manner, forces for molecules with more atoms are inferred using data from smaller-sized molecules, reducing the need for ab initio calculations for on-the-fly training. Cumulative ab initio samples consist of 420, 404, 363, 355, 387, 274, 299, and 200 for alcohol, aldehyde, carboxylate, ester, ether, sugar, lactone, and olide groups, respectively. Similar MD simulations are repeated to generate an independent test set without immediate training. Performance metrics for each expert SGPR ML model on the trained training set are presented in Table 1, using Mean Absolute Error (MAE) and $R^2$.

For SGPR models of alcohol, aldehyde, carboxylate, ester, ether, sugar, lactone, and alkylide groups, a Bayesian Committee Machine (BCM) method is employed to combine them into a single universal model for the smooth estimation of CHO with an arbitrary number of double bonds. The model is reconstructed using both sampled ab initio data and derived descriptors during the training process of ML potential experts for the 8 groups. To assess the efficiency of estimating the values correctly, a sample from the training set is selected, and the estimation efficiency based on the presence or absence of the potential for the group to which the sample belongs is investigated through individual ab initio MD simulations. The results indicate that when the potential for the sample is present in all groups, it requires less than 12 times the energy compared to when it is absent (Table 1).

Regarding the combination of 8 groups into a single universal model, consisting of sets of C, H, O atoms, the estimation efficiency for single universal model was calculated using training sample for each group. As a result of examining the RMSE of 8 groups of samples for a single universal potential, most RMSE of force was lower than 200 meV per angstrom, and RMSE of per-atom energy was lower than 20 meV per atom (Table II). Results of 8 samples for root mean square error (RMSE) of per-atom energy have low scored, despite of the RMSE will always be larger or same score than MAE.

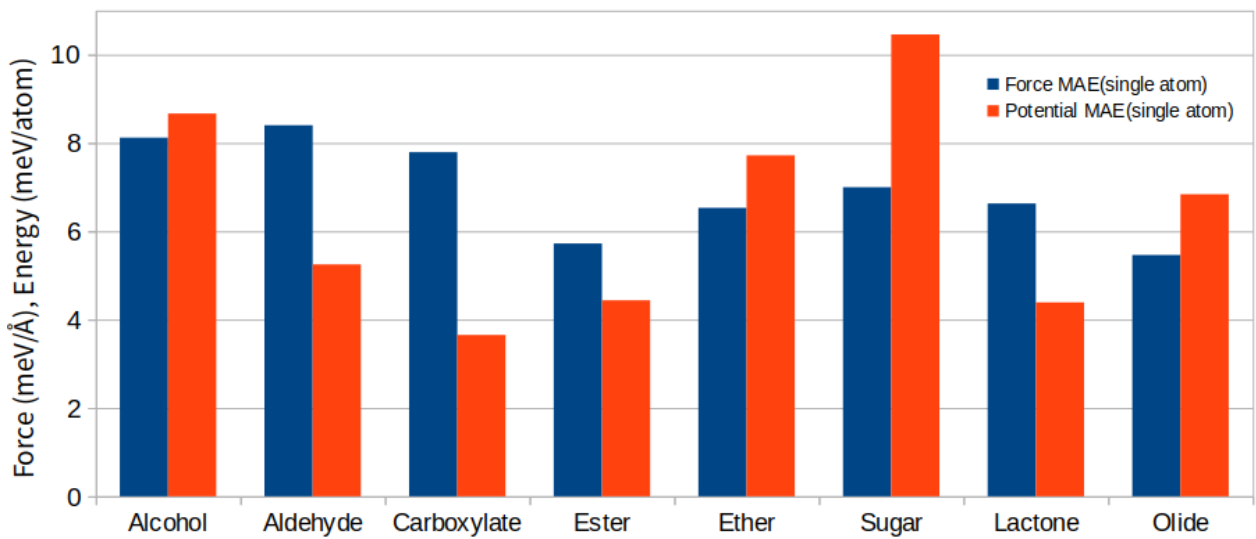

FIG. 3. The average mean absolute errors (MAEs) in test sets were computed for eight groups using the expert SGPR ML potential model.

TABLE II. The root mean square errors (RMSEs) for the single universal model were computed in the training set of the 8 groups.

| Samples | N [a] | RMSE of per-atom energy (meV/atom) | Axis-wise RMSE of Force(meV/Å) |
|---|---|---|---|
| 1,3-Butandiol (Alcohol) | 78 | 12.30296 | 155.71507 |
| Cycloheptanecarbaldehyde (Aldehyde) | 98 | 2.52116 | 106.11747 |
| Pinifolic-acid (Carboxylate) | 89 | 2.88195 | 87.32955 |
| Nuapapuin-a-methyl-ester (Ester) | 81 | 2.49908 | 69.20581 |
| Diethylene-glycol-monoethyl-ether (Ether) | 94 | 6.49553 | 146.20172 |
| D-arabino-hex-2-ulo-pyranose (Sugar) | 69 | 2.99390 | 75.93562 |
| D-glucaro-1,2-lactone (Lactone) | 84 | 4.30684 | 80.91550 |
| Botryolide-D (Olide) | 51 | 8.44353 | 154.69721 |

[a] The number of sample

## DISCUSSION

The CHO molecule exhibits an extremely diverse range of structures, with over 112 million predicted molecules even when considering only 13 carbon atoms[60]. To address this diversity, molecules consisting solely of CHO were organized using internationally agreed-upon IUPAC nomenclature. In this study, molecules containing an arbitrary number of double bonds, cyclo-structures, aromatic rings, and aromatic polycycles within each group were selected and added to the training set (Table S1).

The analysis of structural energy, including structural energy analysis of triple bonds between C and C, and shared bonds between molecules, will be considered in the further research. Handling more complex structures requires further efforts to train MLPs and flexibility to expand the current model.

Additionally, the creation of a system to acquire information, including grid shapes and ion positions for C, H, O molecules, using an automated method or building systems based on canonical smiles and unique atomic information when data is unavailable, poses a significant challenge and requires additional research. These efforts aim to establish an efficient dataset for improved molecular dynamics simulations of CHO molecules.

Using the real-time adaptive sampling and SGPR algorithm, training results for 8 groups yielded MLPs consisting of fewer than 500 atoms each. Furthermore, each MLPs scored under 11 meV per angstrom with potential MAEs for single atom, The energy value learned by each single MLP shows a significantly lower difference in meV per angstrom compared to the DFT energy. Particularly, the BCM method was employed to estimate values for previously trained molecular structures and untrained molecular structures. While BCM considers the weight ratio of MLPs based on the structure of the molecule, providing an advantage in energy estimation, it may occasionally offer slightly lower estimates than well-trained single MLP results.

Whether to check the extent to which the single universal model can be estimated or not, a test was performed selecting a molecule containing a combination of polycycles and double bonds, such as grandiflorenic-acid. As a result of estimation and testing with 100 points, it showed 59.91 meV per atom and 930.97 meV per angstrom. This holds expectation for extending the study to more diverse molecular binding methods and molecules composed of multiple atoms beyond CHO. Further research is planned to investigate the criteria for weight due to Covalent loss when calculating single MLPs through the BCM method.

The combination of SGPR algorithm and BCM method demonstrates a fast and robust approach for on-the-fly machine learning of potential energy surfaces. This combination, coupled with an efficient adaptive sampling algorithm, can generate models consisting of groups of MLPs with hundreds of molecules in the training dataset. Furthermore, the expert model approach provides a modular approach for complex polymer compounds, ultimately leading to the creation of a single universal model.

# ACKNOWLEDGEMENTS

The authors are grateful for computational resources provided by the Korea Institute of Science and Technology Information (KISTI) for the Nurion cluster (KSC-2021-CRE-0542, KSC-2022-CRE-0115). C.W.M. acknowledges the support from the National Research Foundation of Korea (NRF) grant funded by the Korea government (MSIT) (No. NRF-2022R1C1C1010605). SYW and CWM acknowledge the support from the NRF grant RS-2023-00222245. DCY acknowledges the support from the NRF grant RS-2023-00250313.

# A Bayesian Committee Machine Potential for Oxygen-containing Organic Compounds

Seungwon Kim,[1] Changmo David Yang,[1] Soohaeng Yoo Willow,[1, *] and Chang Woo Myung[1, †]

[1]*Department of Energy Science, Sungkyunkwan University, Seobu-ro 2066, Suwon, 16419, Korea*

(Dated: March 2, 2024)

**CONTENTS**

---

[*] sy7willow@skku.edu
[†] cwmyung@skku.edu

TABLE S1. Molecules to use in the training of MLPs

| Group | Molecules |
|---|---|
| Alcohol | Methanol, Ethanol, Ethylene-glycol, 1-Propanol, Isopropyl-alcohol, Propylene-glycol, Allyl-alcohol, 1,3-Propanediol, 2-Methoxyethanol, Glycidol, Isobutanol 1-Pentanol, Cyclopentanol, 3-Pentanol, 2-Pentanol, Piceatannol, Diphenylmethanol, Barakol |
| Aldehyde | Formaldehyde, Acetaldehyde, Propionaldehyde, Malondialdehyde, Crotonaldehyde, Isobutyraldehyde, Cyclobutanecarboxaldehyde, 3-Furaldehyde, Tetrahydrofuran-3-Carbaldehyde, Cyclopentanecarbaldehyde, Paraldehyde, 4-Hydroxybenzaldehyde, Benzaldehyde, 3,4-Dihydroxybenzaldehyde, 3-Hydroxybenzaldehyde, 2,4,6-Trihydroxybenzaldehyde, Cycloheptanecarbaldehyde, o-Phthalaldehyde, 4-Methoxybenzaldehyde, 2-Methoxybenzaldehyde, Phenylacetaldehyde, Metaldehyde, Isophthalaldehyde |
| Carboxylate | Formic-acid, Acetic-acid, Oxalic-acid, Glycolic-acid, Peracetic-acid, Acetic-acid-C-11, Malonic-acid, Propionic-acid, L-Lactic-acid, D-Lactic-acid, Lactic-acid, Acrylic-acid, Pinifolic-acid, Boropinic-acid |
| Ester | Methyl-formate, Methyl-acetate, Methyl-acrylate, Methyl-butyrate, Propyl-acetate, Ethyl-propiolate, Ethyl-acrylate, Butyl-acetate, Isobutyl-acetate, Ethyl-isobutyrate, Ethyl-butyrate, Nuapapuin-a-methyl-ester |
| Ether | Diethyl-ether, Diethylene-glycol, Vinyl-ether, Diethylene-glycol-monoethyl-ether, 2-Methoxycinnamic-acid, 2',4'-Dimethoxyacetophenone, 7-Methoxy-4-methylcoumarin, Diphenyl-ether, 2,5-Dimethoxy-p-cymene, 3,4,5-Trimethoxycinnamic-acid, Tetrahydrosapponone-a-trimethyl-ether |
| Sugar | $\alpha$-D-Ribofuranose, Arabinose, D-Xylitol, DL-Arabinose, $\beta$-D-Ribopyranose, D-Lyxopyranose, D-Arabino-Hex-2-ulo-Pyranose, 6-Deoxy-Galactopyranose, D-Mannose, 2,6-Dideoxy-D-Ribo-Hexopyranose |
| Lactone | D-Glucaro-1,4-lactone, D-Glucurono-3,6-lactone |
| Olide | Botryolide-D, Botryolide-B, Botryolide-A, Stagonolide-E, Stagonolide-F, Stagonolide-D, Stagonolide-C |

**S0.1. Molecules to use in the training of MLPs**